\documentclass{svproc}

\usepackage{url}
\usepackage{amsmath}
\usepackage{graphicx}
\usepackage[strings]{underscore}
\usepackage{subcaption}
\begin{document}
\mainmatter              
\title{Impact of network topology on efficiency of proximity measures for community detection}
\titlerunning{Impact of network topology on measures efficiency for community detection}  
%
\author{Rinat Aynulin}

\authorrunning{Rinat Aynulin} 
%
\tocauthor{Rinat Aynulin}

\institute{Kotel’nikov Institute of Radio-engineering and Electronics (IRE) of Russian Academy of
Sciences, Mokhovaya 11-7, Moscow 125009, Russia \\
Moscow Institute of Physics and Technology, 9 Inststitutskii per., Dolgoprudny, Moscow region, 141700 Russia, \email{rinat.aynulin@phystech.edu}}
\maketitle              

\begin{abstract}
Many community detection algorithms require the introduction of a measure on the set of nodes. Previously, a lot of efforts have been made to find the top-performing measures. In most cases, experiments were conducted on several datasets or random graphs. However, graphs representing real systems can be completely different in topology: the difference can be in the size of the network, the structure of clusters, the distribution of degrees, the density of edges, and so on. Therefore, it is necessary to explicitly check whether the advantage of one measure over another is preserved for different network topologies. In this paper, we consider the efficiency of several proximity measures for clustering networks with different structures. The results show that the efficiency of measures really depends on the network topology in some cases. However, it is possible to find measures that behave well for most topologies.

\end{abstract}

\let\thefootnote\relax\footnotetext{This is a pre-copyedited version of a contribution published in Studies in Computational Intelligence edited by Cherifi H., Gaito S., Mendes J., Moro E., Rocha L., published by Springer, Cham. The definitive authenticated version is available online via https://doi.org/10.1007/978-3-030-36687-2_16}

\section{Introduction}
Many systems from biology, chemistry, computer science, social science, etc. can be represented as networks \cite{applications}. One of the fundamental features of such networks is community structure or clustering. Usually, a cluster refers to a group of nodes that have more edges to each other than to members of other clusters.

Community detection is a popular topic, and a lot of methods have been proposed to solve this problem, some of which use the notion of a proximity (or distance) measure on the set of graph nodes. For a long time, mathematicians used only the shortest path distance as a measure, however, later more complex measures were introduced \cite{DezaDeza16}. Choice of measures can improve or worsen results of graph clustering, so it is important to find the most efficient measures.

As already noted, real-world networks can represent different systems from different domains. Although they may have some common properties (e.g., power law degree distribution or scale-free property), it would be wrong to expect that they are all the same in structure. Indeed, the research results show a profound difference in the topological properties of networks \cite{structure-functions}.

Previously, it was shown that the efficiency of community detection methods significantly depends on the network topology \cite{pasta-topology}. The main goal of this paper is to check if it is true for different proximity measures within the framework of a fixed clustering algorithm. We consider several proximity measures, i.e., Walk, Communicability, Forest, Heat, and PageRank. Their efficiency is tested in experiments in which networks with different topologies generated using the LFR model are clustered using the Spectral and the Ward method.

According to the results of the experiments, there is some relation between the efficiency of proximity measures and network topology. However, in most cases, the ranking of measures by the quality doesn't change, and we can find the top-performing measures.

\section{Related Work}
\label{sec:relwork}
This section is divided into 2 parts. First, we discuss studies in which different measures are compared. In the second part, we provide a brief overview of previous works which take into account network topology when discussing clustering.

In \cite{comparison-felix}, authors used the Randomized Shortest-Path, Free Energy, Sigmoid Commute-Time, Corrected Commute-Time, and Logarithmic Forest measure for community detection in 15 real datasets. The top-performing measures were Randomized Shortest-Path and Free Energy. In \cite{comparison-documents}, 5 graph metrics are compared for document collections clustering.

\cite {comparison-ect} shows the superiority of the Euclidean Commute Time metric over the standard Euclidean Distance. Measures are tested in experiments with several artificial datasets.

The concept of transformation of proximity measure was proposed in \cite{comparison-logarithmic,transformations}, and experiments on random graphs and some classical datasets revealed an increase in clustering quality when using transformed measures.

In \cite{cheb-on-kernels}, a number of proximity measures, including Walk, Communicability, Heat, PageRank, and several logarithmic measures are used to find communities in SBM (stochastic block model) based networks with the Spectral method, and some of the measures lead to better results than others.

Previously, network topology has been already considered in the context of community detection. In \cite{at-scale}, the performance of such algorithms as Louvain, Infomap, label propagation, and smart local moving is examined for synthetic graphs and empirical datasets varying in size and edge density. In \cite{think-locally}, authors investigate the relationship between clusters quality and their sizes.

\cite{pasta-topology} provides a study of the performance of community detection algorithms when applied to networks with different topologies and reveals some limitations of the current methods.

\section{Background and Preliminaries}
\subsection{Definitions}
We consider an undirected graph $G = (V, E)$, where $V$ is the set of nodes and $E$ is the set of edges (i.e., 2-element subsets of $V$). The \textit{degree} of a node is the number of edges connected to it, and the degree matrix $D = \mathrm{diag} (A \cdot \textbf{1})$ contains information about nodes' degrees. The adjacency matrix $A = (a_{ij})$ for a graph $G$ is a square matrix with $a_{ij} = 1$ if there is an edge from node $i$ to node $j$, and $a_{ij} = 0$ otherwise. The Laplacian matrix $L$ is defined as $L = D - A$, and $P = D^{-1}A$ is the Markov matrix.

Given a graph $G$, by a \textit{measure} we mean a function $\kappa$ on the set of pairs of nodes that characterizes proximity or similarity between graph nodes. A \textit{kernel} on a graph is a similarity measure which can be represented as a Gram matrix, i.e., symmetric positive semidefinite matrix $K$ \cite{cheb-on-kernels}.

\subsection{Clustering methods}
In this paper, we study 2 clustering methods: Ward and Spectral.

\subsubsection{Ward.}

This method was proposed by Ward in 1963 \cite {ward}. The Ward method is hierarchical agglomerative. The idea behind it is that at the first step of the algorithm each object is considered as a separate cluster. In the subsequent steps, the closest clusters are combined. 

The distance between two clusters is defined as $\delta(A, B) = \sum_{x_i \in A \cup B} d^2(x_i - m(A \cup B)) - \sum_{x_i \in A } d^2(x_i - m(A)) - \sum_{x_i \in B} d^2(x_i - m(B))$, where $m(A)$ is the center of cluster $A$. So, it is the increase in the sum of the squares of the distances of nodes to the cluster centers in the case of combining these clusters.

Initially, the sum of the squares is zero: each node is in its cluster. Then the clusters are combined so that the increment of the sum of squares is minimal.

\subsubsection{Spectral.}
The general approach to implement the Spectral clustering is to apply the $k$-Means or another standard clustering method to eigenvectors of the Laplacian matrix of the graph. 

For a detailed review and intuition behind the Spectral algorithm, we refer to an excellent tutorial by Ulrike von Luxburg \cite{spectral-tutorial}.

\subsection{Measures}
\label{sec:measures}
We consider the following measures:

\begin{itemize}
    \item Walk: $K = \sum_{n=0}^{\infty} \alpha ^ n A ^ n = (I - \alpha A) ^ {-1}$, $\alpha \in (0, q^{-1}) $, where $q$ is the spectral radius of the adjacency matrix of a graph \cite{cheb-walk,walk}
    \item Communicability (Comm): $K = \sum_{n=0}^{\infty} \frac{ \alpha ^ n A ^ n}{n!}= \mathrm{exp}(\alpha A)$, $\alpha > 0$ \cite{comm-distance,comm-distance-2}
    \item Forest: $K =  \sum_{n=0}^{\infty} \alpha ^ n (-L) ^ n  = (I + \alpha L) ^ {-1}$, $\alpha > 0$ \cite{cheb-forest-kernel}
    \item Heat: $K = \sum_{n=0}^{\infty} \frac{ \alpha ^ n (-L) ^ n}{n!}= \mathrm{exp}(-\alpha L)$, $\alpha > 0$ \cite{heat-kernel}
    \item PageRank (PR): $K = (I - \alpha P) ^ {-1}$, $0 < \alpha < 1$ \cite{pagerank}
\end{itemize}

It is important to note that because of the kernels definitions, the eigenvectors are the same for the Walk and Communicability, and Forest and Heat measures \cite{cheb-on-kernels}. Consequently, the Spectral clustering will lead to the same partitions for these 2 pairs, and we will use only 3 measures (i.e., Walk, Forest, and PageRank) instead of 5 when discussing results for the Spectral method.
\subsection{Network Generation}
To generate networks with different topologies, we use the LFR model introduced by Lancichinetti et al. in \cite{LFR}. Generated networks share a number of features which real networks have, e.g., the power law degree and community size distributions. 

Changing the model input parameters, one can get networks varying in size, average degree, power law exponent for the degree and community size distributions, minimum and maximum size of clusters, and clusters quality, i.e., the fraction of inter-community edges. Together, this allows one to get graphs with completely different structures. We discuss chosen input parameters and reasons for this choice in Section \ref{sec:method}.

\subsection{Clustering Quality Evaluation}
For clustering quality evaluation, the Adjusted Rand Index (ARI) introduced in \cite{ari-hubert} is used. Reasons for using this quality index are provided in \cite{ari-best}.

ARI plays an important role in the study, so we will briefly explain it.

Initially, the Rand Index was introduced in \cite{rand}. If $X$ and $Y$ are two different partitions (clusterings) of $n$ elements, let $a$ be the number of pairs of elements that are in the same clusters in $X$ and $Y$, and $b$ the number of pairs of elements that are in different clusters in $X$ and $Y$. Then the Rand Index equals to $\frac{a + b}{\binom{n}{2}}$. 

The idea here is simple: it is the number of agreements between two partitions divided by the total number of pairs. Unfortunately, the Rand Index has a drawback: the expected value of the Rand Index is not zero for random partitions. So, it should be corrected, and ARI is the corrected version of the Rand Index: $ARI = \frac{\rm{Index} - \rm{ExpectedIndex}}{\rm{MaxIndex} - \rm{ExpectedIndex}}$.

For ARI, $1$ refers to perfect matching, while $0$ is characteristic of random labeling.

\section{Experimental Methodology}
In this study, measures are tested in experiments with networks generated using the LFR model. To obtain different topologies, the following 6 input parameters of the LFR model are varied:
\begin{itemize}
    \item \textbf{Network size ($n$)}. 
    Obviously, real networks are different in size. Due to the computational limits, we cannot generate really big networks, so graphs are generated with the following numbers of nodes: $100, 300, 500, 1000, 2000, 3000$.
    \item \textbf{Average degree ($m$)}.
    Varying average degree from $2$ to $15$ with step $1$ allows obtaining networks from very sparse (for which the clustering quality will be close to 0, regardless of other parameters) to pretty dense (with the clustering quality close to 1 for networks with good community structure).
    \item \textbf{Power law exponent for the degree distribution ($\tau_1$)}.
    The power law exponent is usually considered to be in the range from $2$ to $3$ \cite{structure-functions,LFR}. We use the following values of $\tau_1$: $2.0, 2.2, 2.4, 2.5, 2.6, 2.8, 3.0$.
    \item \textbf{Power law exponent for the community size distribution ($\tau_2$)}. Like the degree distribution, the community size distribution was also reported to follow the power law with the typical limits $1 < \tau_2 < 2$ \cite{LFR}. Networks generated in this study have the power law exponent for the community size distribution varying from $1$ to $2$ with step $0.25$.
    \item \textbf{Minimum and maximum communities size ($\rm{cmin}$ and $\rm{cmax}$)}. Changing the limits for the communities size, we can get networks with a lot of small communities, few big communities, and intermediate stages between them. As a baseline, for $n = 300$, the following limits are used: $[20, 50], [50, 80], [80, 140], [140, 185]$, and if the network size is different, then the limits are scaled accordingly.
    \item \textbf{Fraction of inter-community edges ($\mu$)}. This parameter allows to change the quality of communities. We vary $\mu$ in the range from $0.1$ to $0.6$ with step $0.1$.
    
Graphs are generated in the following way: the basic configuration is $n = 300$, $m = 5$, $\tau_1 = 2.5$, $\tau_2 = 1.5$, $cmin = 80$, $cmax = 140$, $\mu=0.2$. Then, one of the parameters varies from the basic configuration within the limits described above.
After generation, networks are clustered with each of the measures listed in Section \ref{sec:measures} using the Ward and Spectral method. 

Each of the measures depends on the parameter. Therefore, we also search for the optimal parameter, and the results include the clustering quality for the optimal parameter. As already noted, the quality of produced partitions is evaluated using ARI.
To get a stable result, for each combination of parameters, 100 graphs are generated and the quality index is averaged over them.
\end{itemize}
\label{sec:method}
\section{Results}
In this section, we discuss the results of the experiments described above.

\begin{figure}[t!]
\begin{subfigure}{0.48\textwidth}
\includegraphics[width=\linewidth]{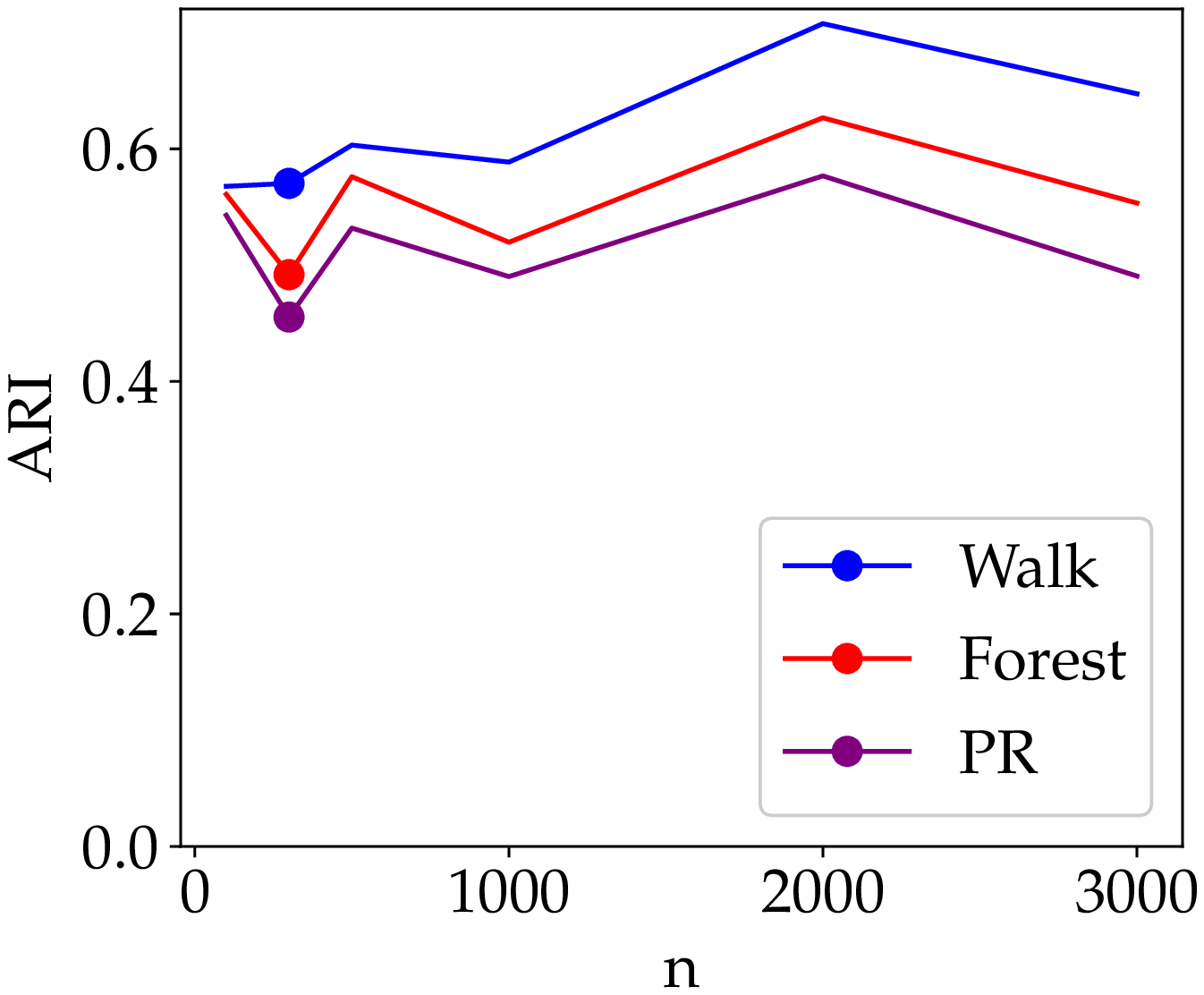}
\caption{The size of the network is changing} 
\label{fig:spectral-all-a}
\end{subfigure}\hspace*{\fill}
\begin{subfigure}{0.48\textwidth}
\includegraphics[width=\linewidth]{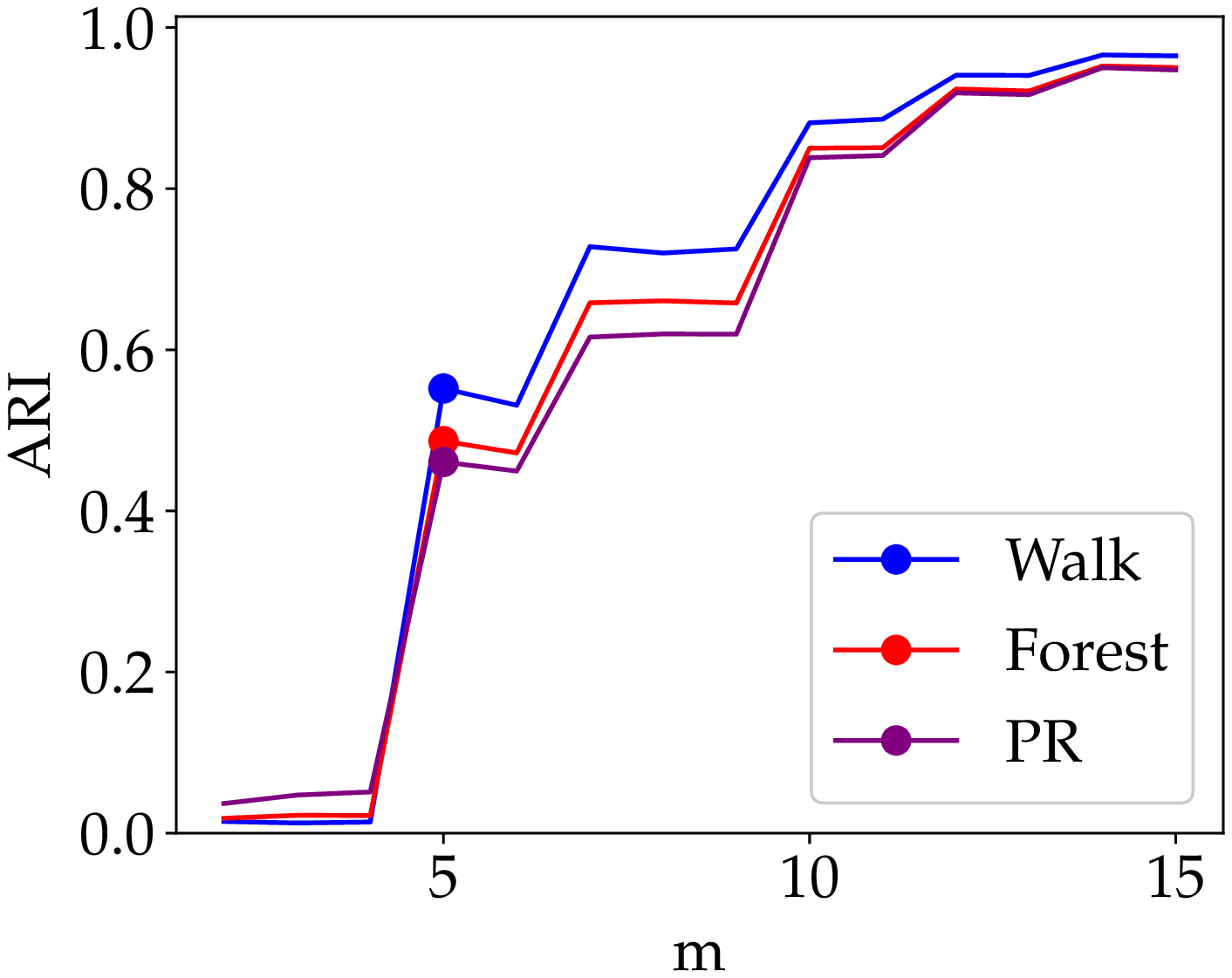}
\caption{The average degree is changing}
\label{fig:spectral-all-b}
\end{subfigure}
\smallskip
\begin{subfigure}{0.48\textwidth}
\includegraphics[width=\linewidth]{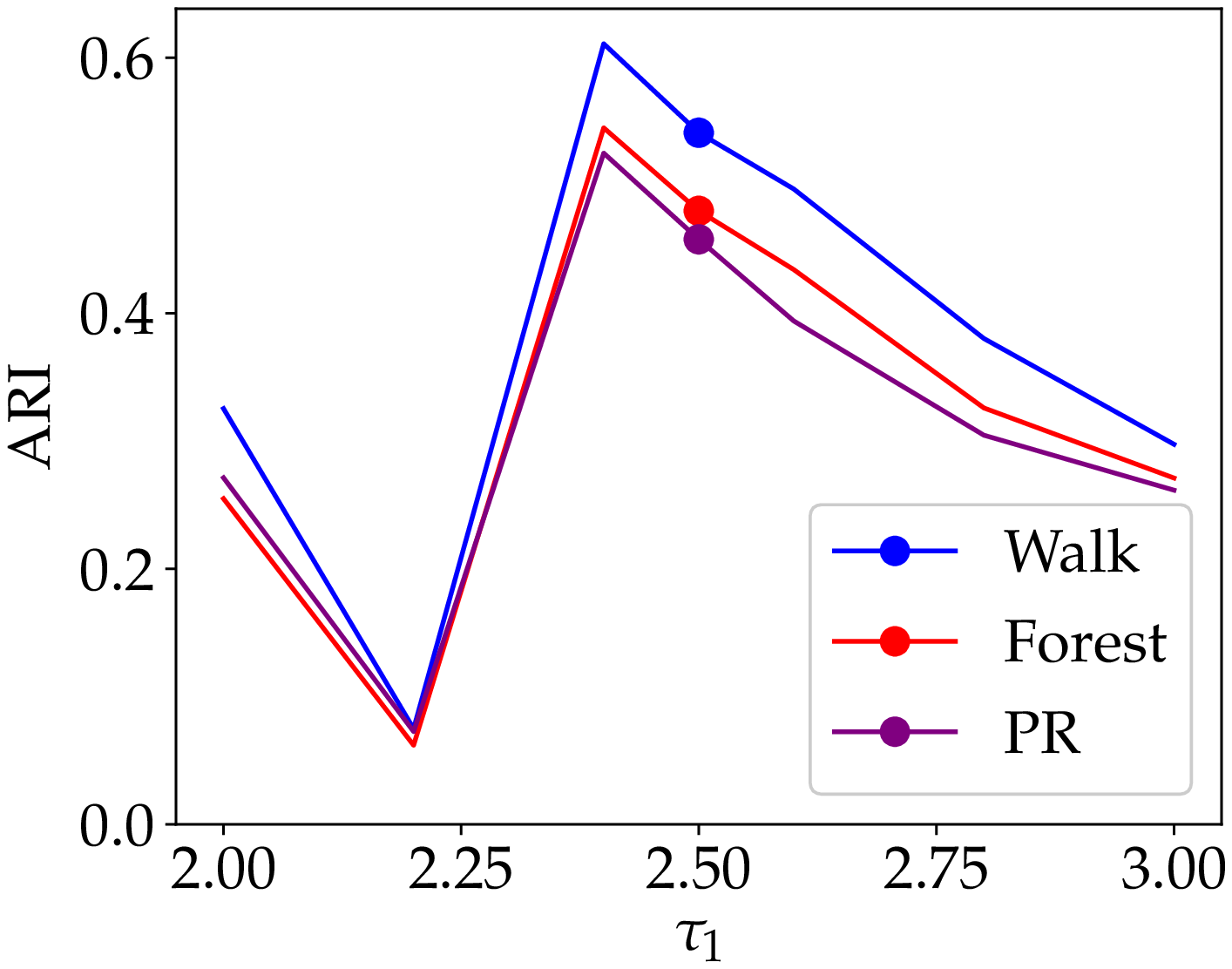}
\caption{The power law exponent for the degree distribution is changing} 
\label{fig:spectral-all-c}
\end{subfigure}\hspace*{\fill}
\begin{subfigure}{0.48\textwidth}
\includegraphics[width=\linewidth]{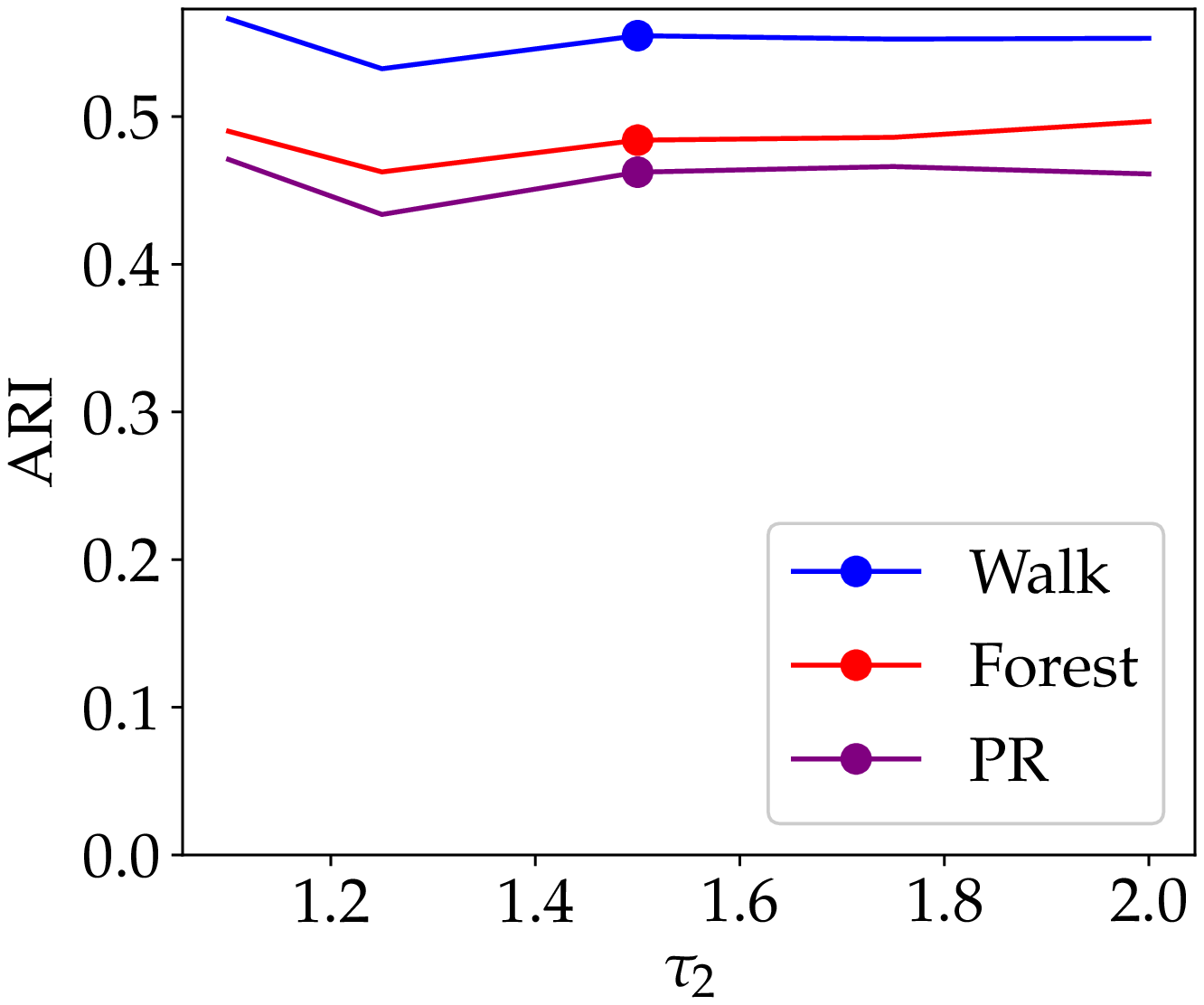}
\caption{The power law exponent for the community size distribution is changing} 
\label{fig:spectral-all-d}
\end{subfigure}

\smallskip
\begin{subfigure}{0.48\textwidth}
\includegraphics[width=\linewidth]{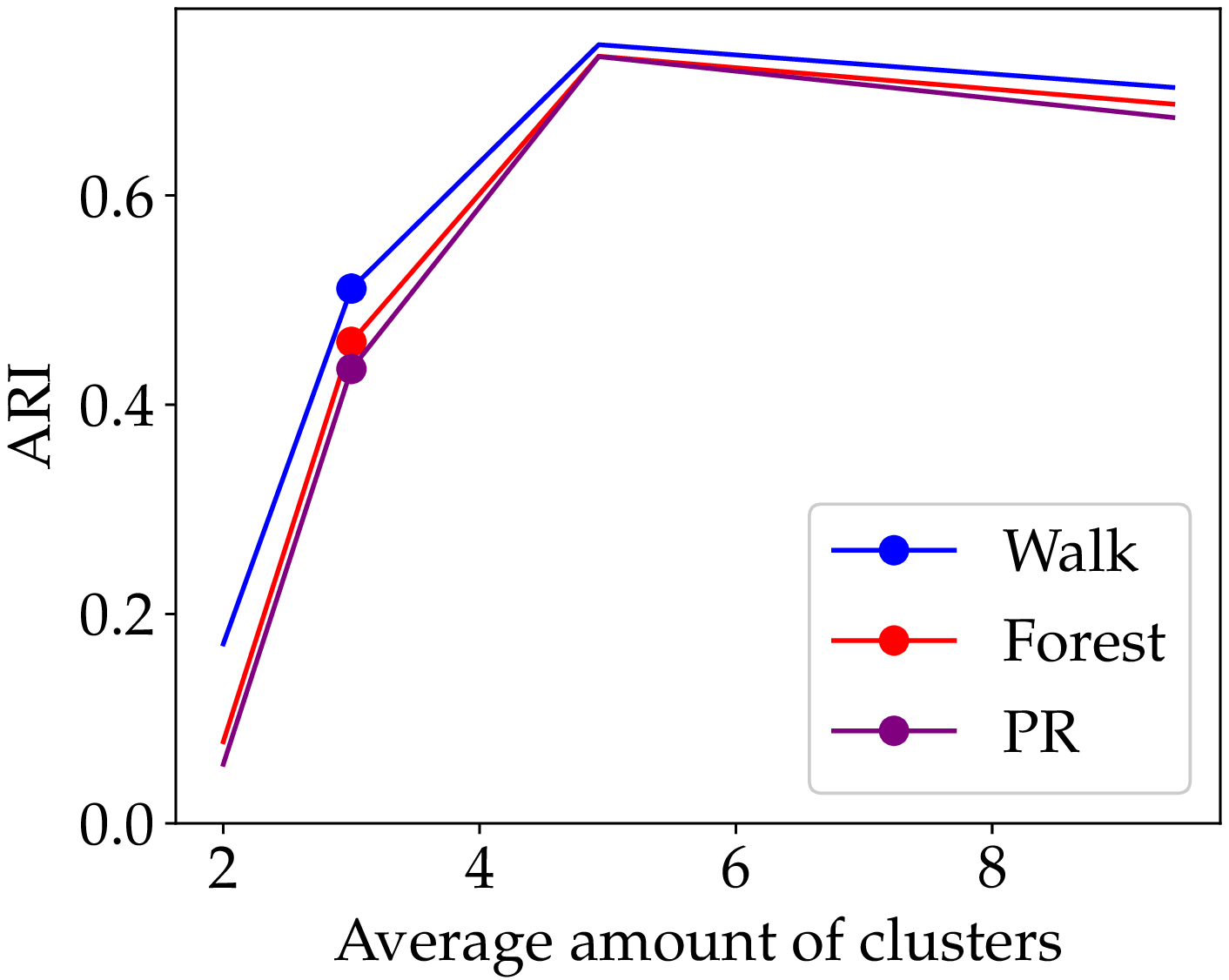}
\caption{The cluster size limits are changing} 
\label{fig:spectral-all:e}
\end{subfigure}\hspace*{\fill}
\begin{subfigure}{0.48\textwidth}
\includegraphics[width=\linewidth]{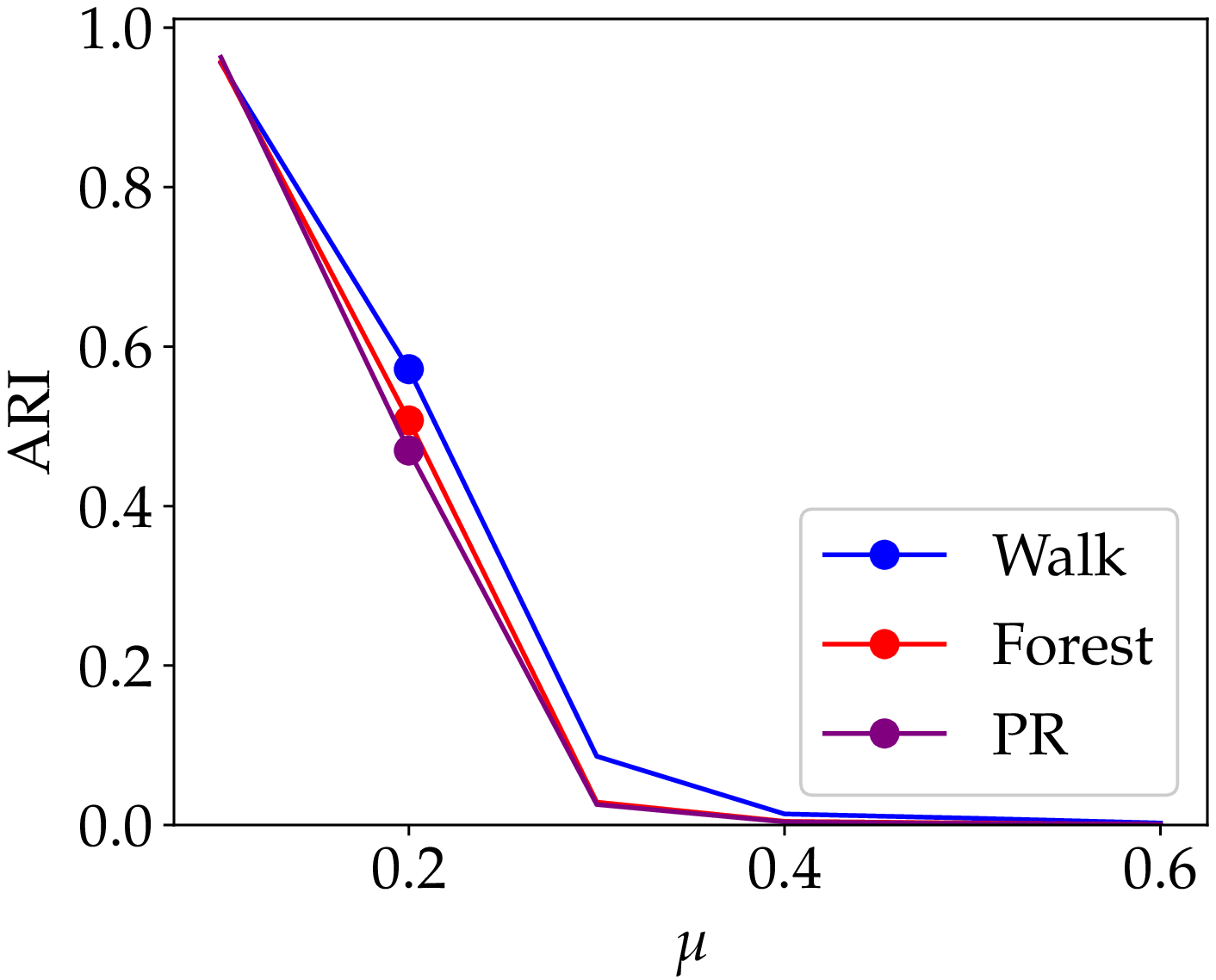}
\caption{The fraction of inter-cluster edges is changing} 
\label{fig:spectral-all-f}
\end{subfigure}
\caption{Results for the Spectral method, point $n = 300$, $m = 5$, $\tau_1 = 2.5$, $\tau_2 = 1.5$, $cmin = 80$, $cmax = 140$, $\mu = 0.2$ is marked} 
\label{fig:spectral-all}
\end{figure}

Figure \ref{fig:spectral-all} presents results for the Spectral method. As noted in Section \ref{sec:measures}, it is meaningful here to analyze results only for 3 measures out of 5 under research. The basic set of parameters is marked by a circle on each of the graphs. The $x$-axis shows the values of the varying parameter, while the value of average ARI is plotted on $y$-axis. When changing the community size limits, the average number of clusters for the generated networks is plotted on the $x$-axis.

Both the clustering algorithm and the proximity measure may depend on the network topology. Common features of the plots for different proximity measures show how the algorithm depends, while deviation from the general picture shows the dependence of the proximity measure on the network topology.

As can be seen, for the Spectral algorithm, all proximity measures behave similarly when topology changes. Their ranking by the quality index is also largely preserved. So we can conclude that when using the Spectral method, the dependence of the relative clustering quality on the network topology for each measure is almost absent, and the top-performing measure is Walk for most topologies.

Let's now look at common features of the plots for different measures and analyze how the clustering quality depends on the network structure for the Spectral algorithm itself.

According to Figure \ref{fig:spectral-all-a}, the Spectral method copes well with the network size increase. This is not generally true for all community detection algorithms \cite{pasta-topology}.

There is a rapid increase in the clustering quality when the average degree increases (Figure \ref{fig:spectral-all-b}). This can be explained by the definition of the community on which the Spectral method bases. Like most clustering algorithms, this method looks for groups of nodes which are densely connected, and it is hard to do it when there are almost no edges in the network.

Figure \ref{fig:spectral-all-c} reveals an interesting relationship between the quality and the power law exponent for the degree distribution. For example, there are several local maxima and minima, and after the local minimum at $\tau_1 = 2.2$ there is the peak at $\tau_1 = 2.4$. So far, there is no explanation for such behavior, and it can be explored more in-depth in future studies.

In Figure \ref{fig:spectral-all-d}, we can see that the quality is almost independent of $\tau_2$. According to Figure \ref{fig:spectral-all:e}, the clustering quality is better when there are a lot of small communities than there are few big communities. 

Finally, in Figure \ref{fig:spectral-all-f}, one can see an expected steep decline in the quality when the fraction of inter-cluster edges increases.

\begin{figure}[t!]
\begin{subfigure}{0.48\textwidth}
\includegraphics[width=\linewidth]{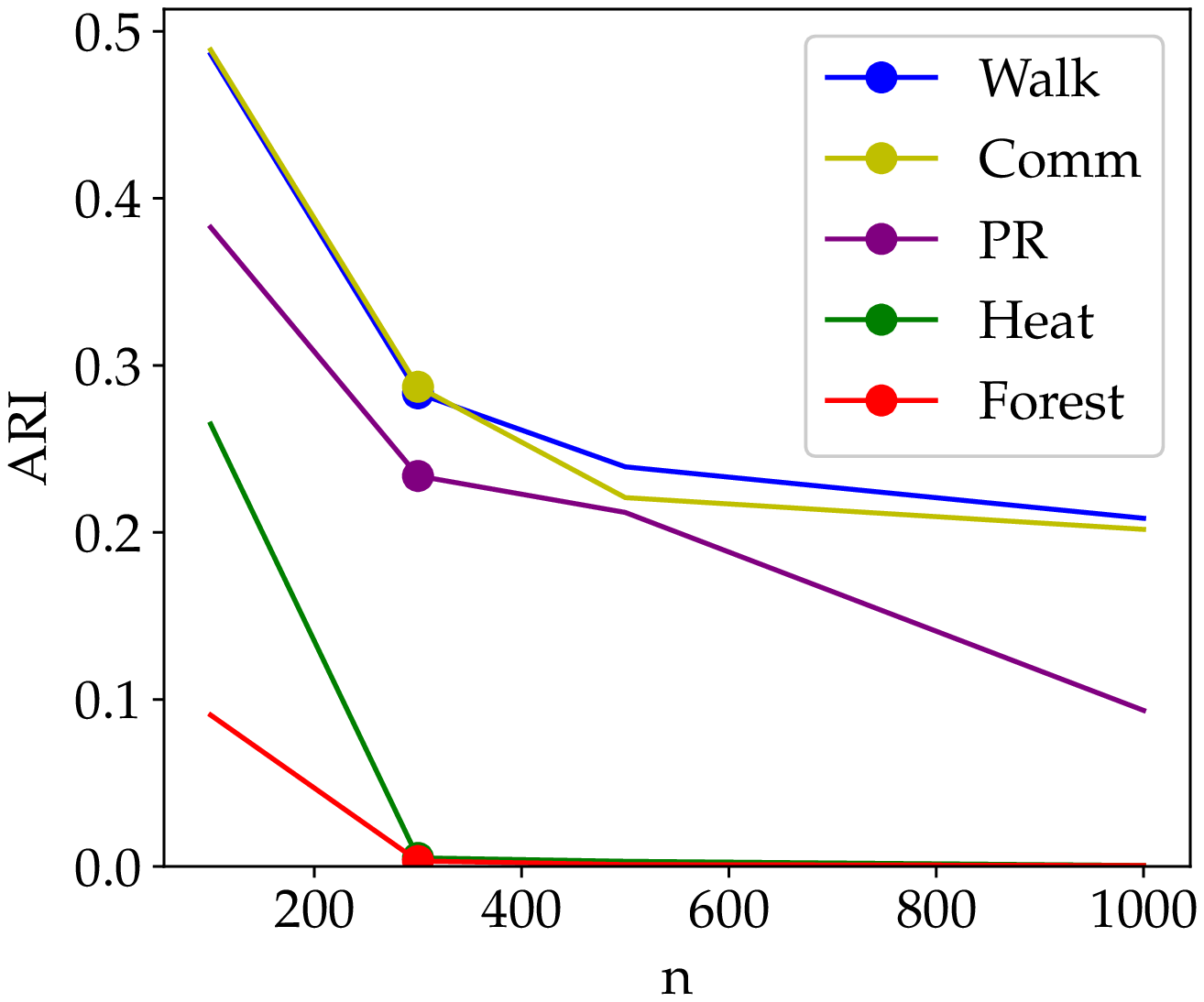}
\caption{The size of the network is changing} 
\label{fig:ward-all-a}
\end{subfigure}\hspace*{\fill}
\begin{subfigure}{0.48\textwidth}
\includegraphics[width=\linewidth]{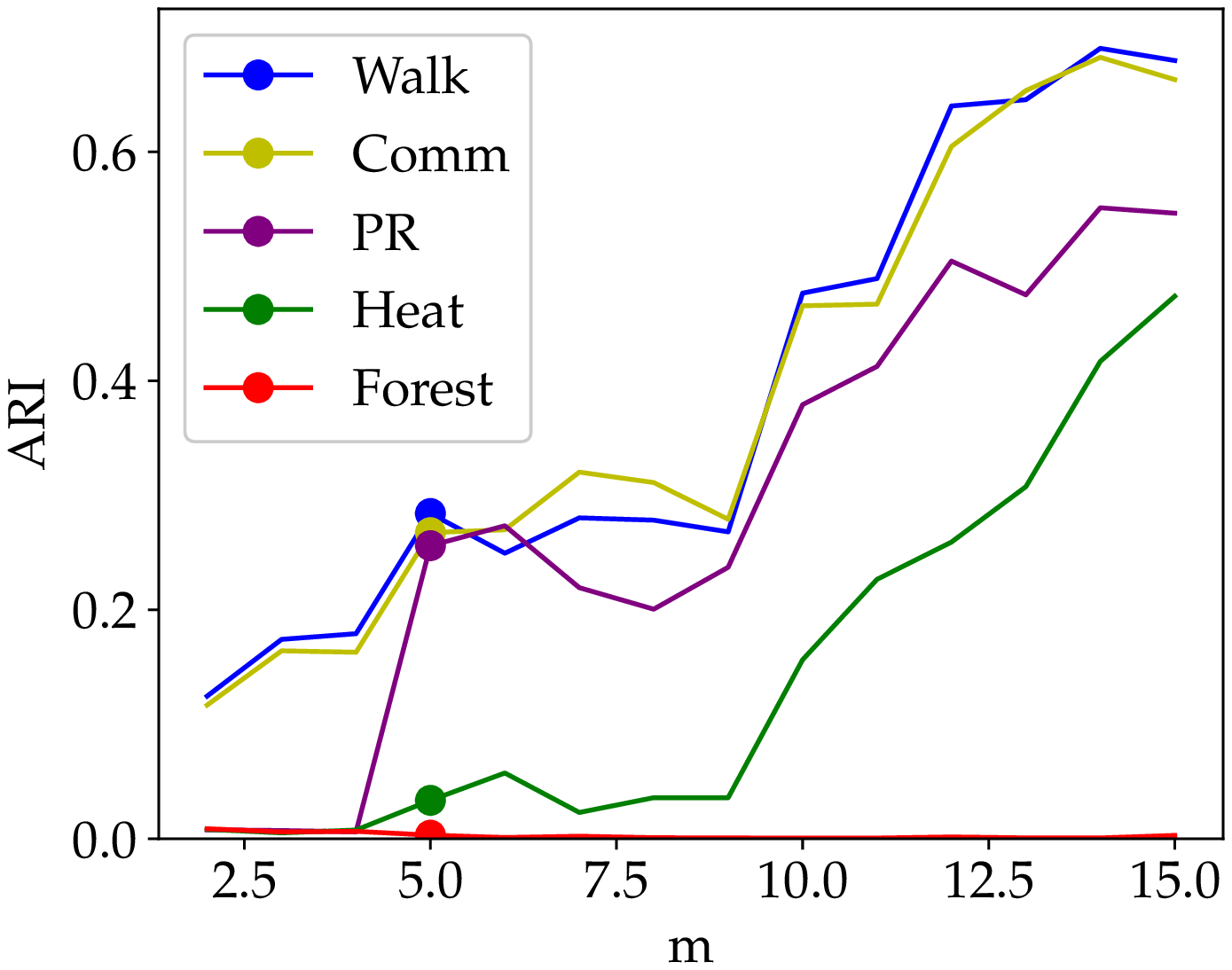}
\caption{The average degree is changing}
 \label{fig:ward-all-b}
\end{subfigure}

\smallskip
\begin{subfigure}{0.48\textwidth}
\includegraphics[width=\linewidth]{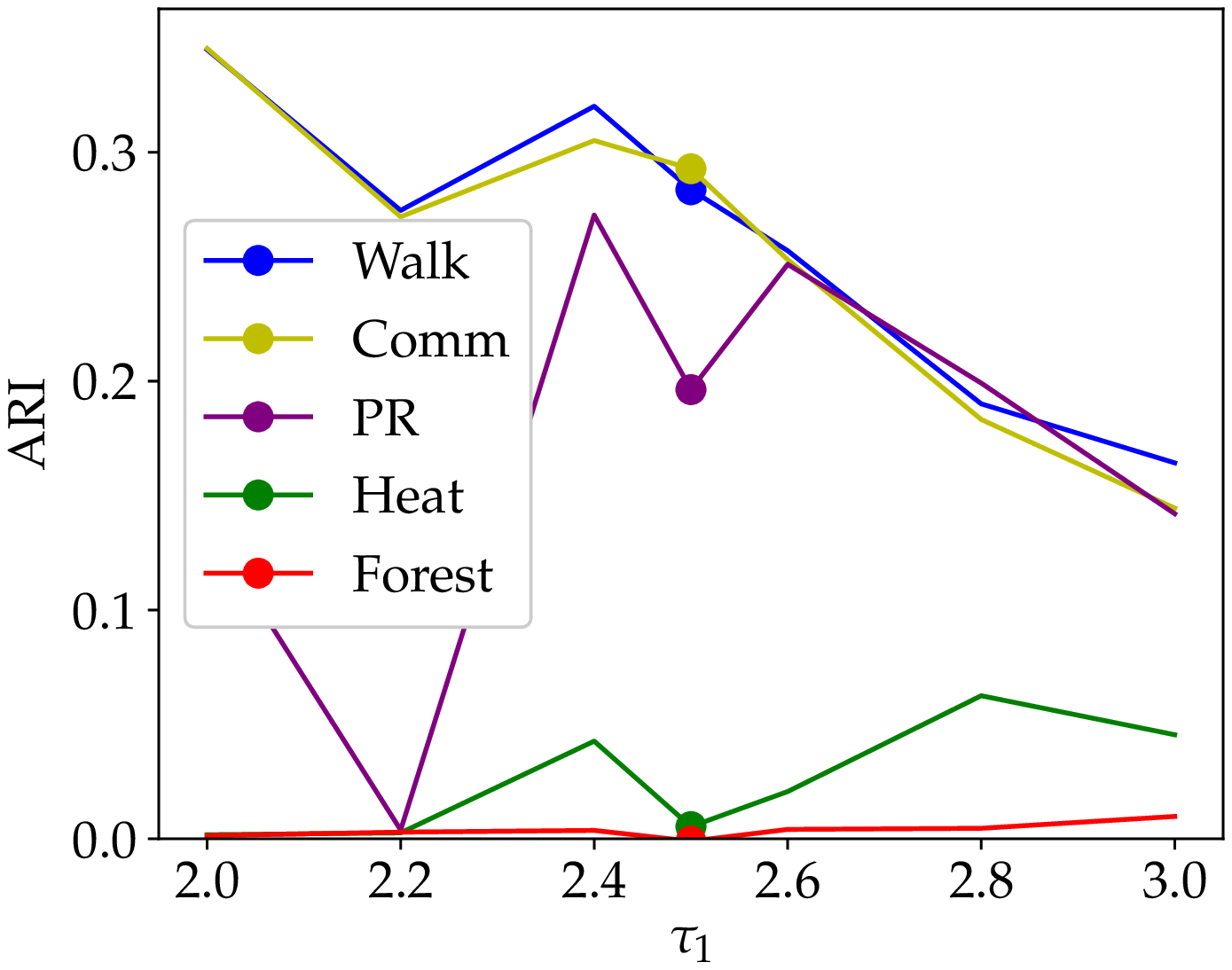}
\caption{The power law exponent for the degree distribution is changing} 
\label{fig:ward-all-c}
\end{subfigure}\hspace*{\fill}
\begin{subfigure}{0.48\textwidth}
\includegraphics[width=\linewidth]{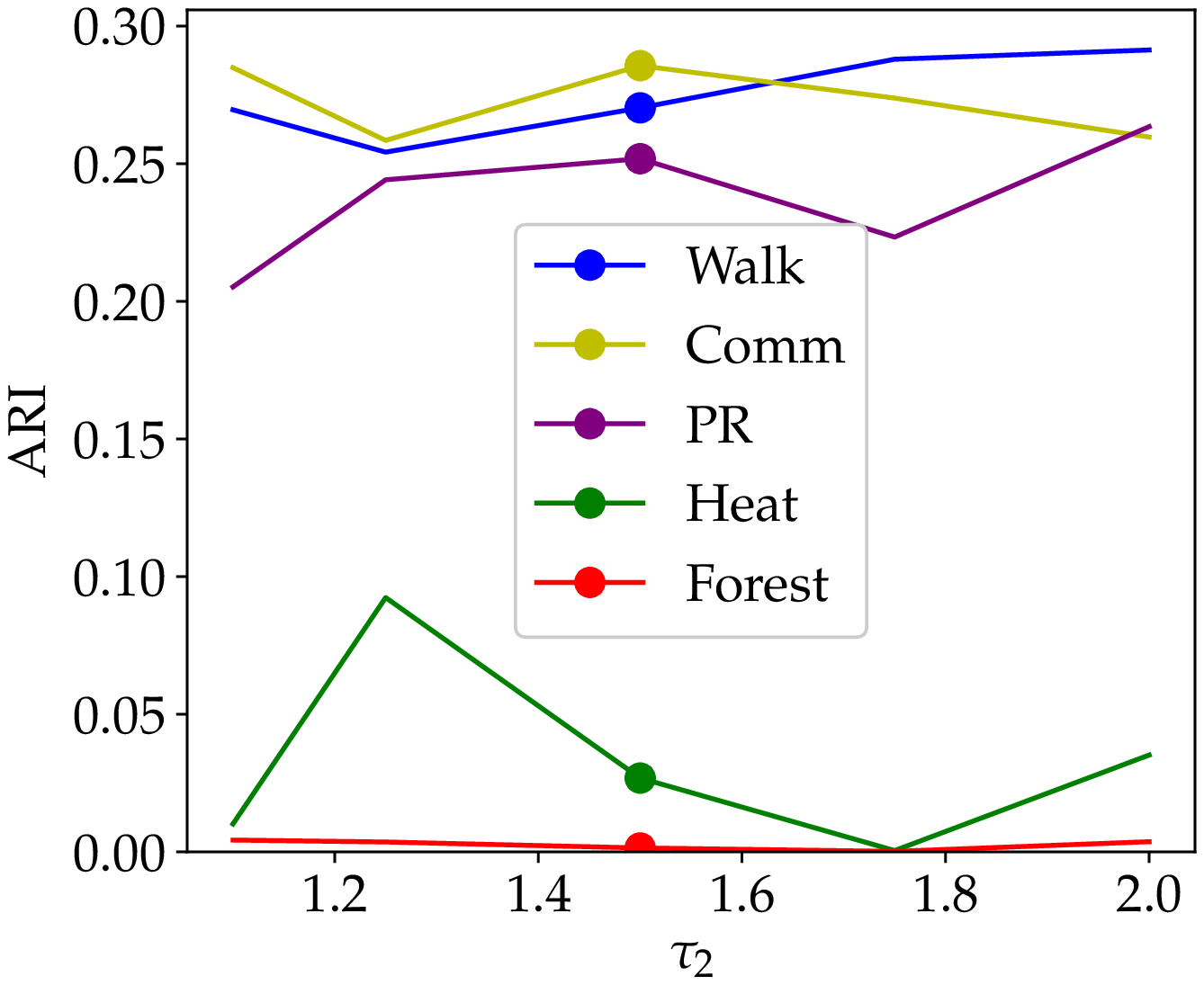}
\caption{The power law exponent for the community size distribution is changing} 
\label{fig:ward-all-d}
\end{subfigure}

\smallskip
\begin{subfigure}{0.48\textwidth}
\includegraphics[width=\linewidth]{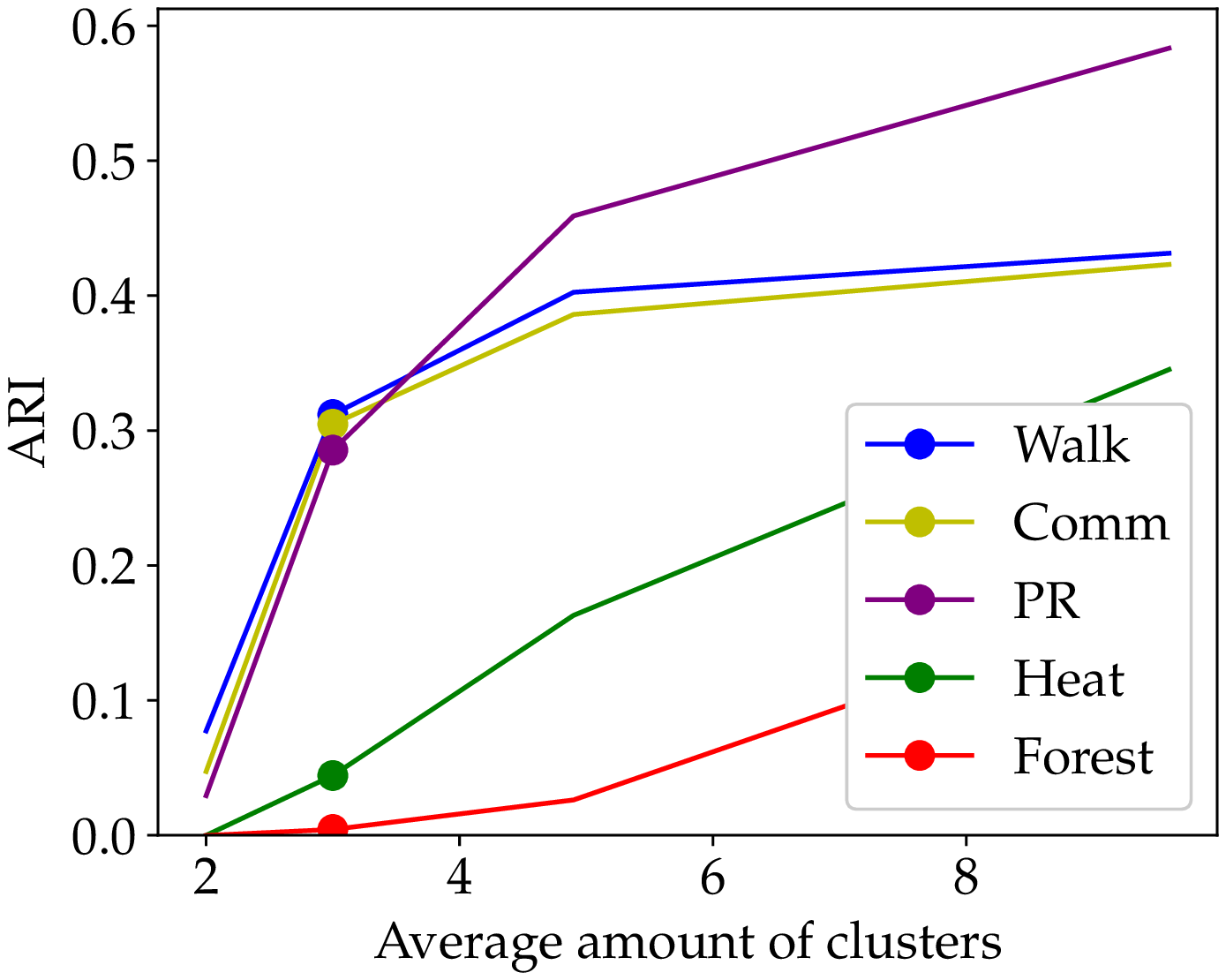}
\caption{The cluster size limits are changing} 
\label{fig:ward-all:e}
\end{subfigure}\hspace*{\fill}
\begin{subfigure}{0.48\textwidth}
\includegraphics[width=\linewidth]{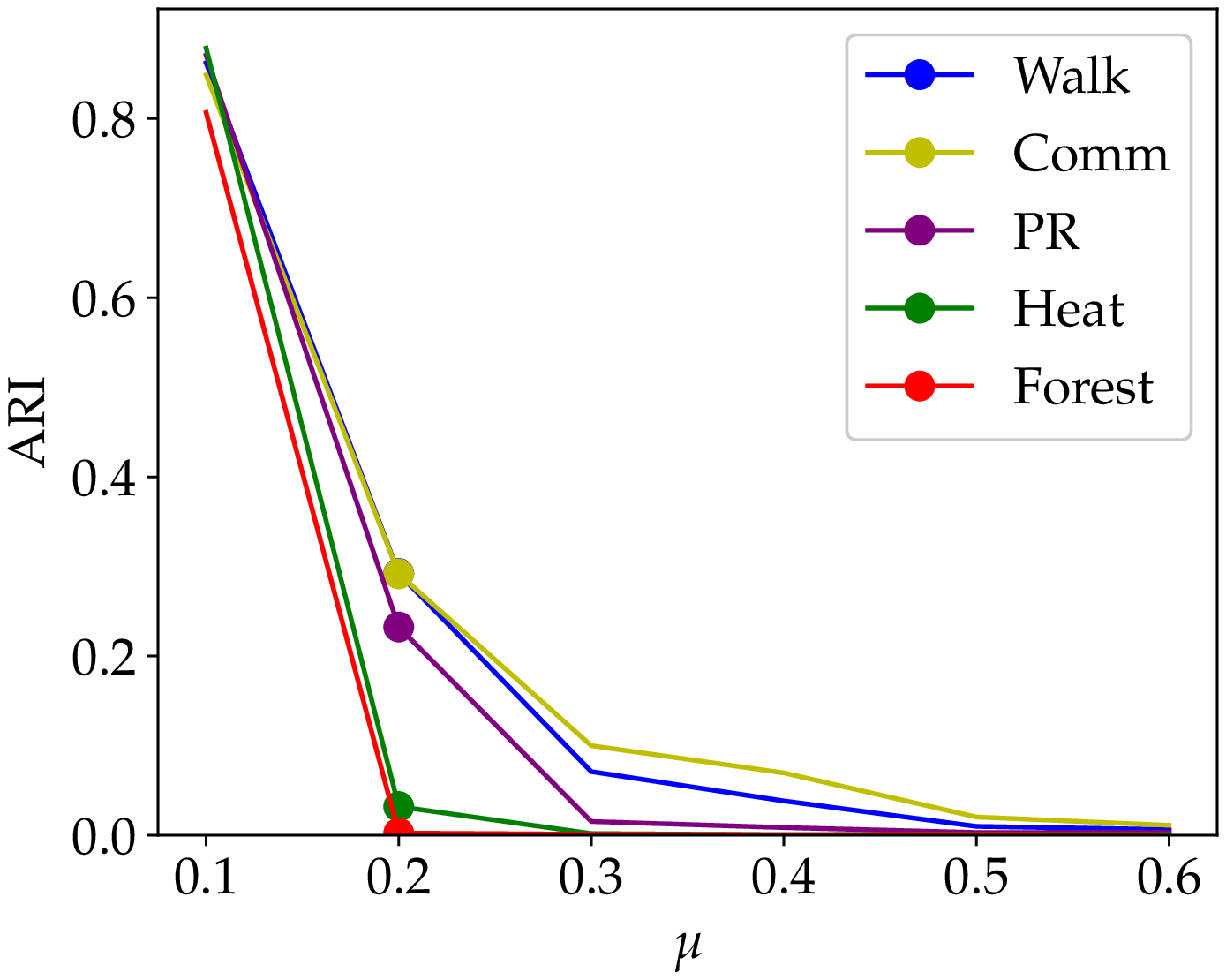}
\caption{The fraction of inter-cluster edges is changing} 
\label{fig:ward-all-f}
\end{subfigure}
\caption{Results for the Ward method, point $n = 300$, $m = 5$, $\tau_1 = 2.5$, $\tau_2 = 1.5$, $cmin = 80$, $cmax = 140$, $\mu = 0.2$ is marked} 
\label{fig:ward-all}
\end{figure}

Results for the Ward algorithm are presented in Figure \ref{fig:ward-all}. This algorithm is more sensitive to the choice of the proximity measure. This is most noticeable in Figure \ref{fig:ward-all-c}, when we vary the power law exponent for the degree distribution. However, we can still find the measures which perform well for most topologies (Walk and Communicability), and the ranking of measures by the quality generally remains the same. So, generally the superiority of one measure over another is the fundamental property which doesn't depend on the network topology. However, there are a few exceptions. For example, PageRank outperforms all the other measures when there are a lot of small clusters (Figure \ref{fig:ward-all:e}).

An interesting relation can be seen in Figure \ref{fig:ward-all-f}. According to it, the Forest and Heat measures are slightly worse than others when there are clear cluster structure and $\mu = 0.1$. But as soon as the cluster structure becomes slightly less distinct, and the fraction of inter-cluster edges increases to $0.2$, their quality drops rapidly to zero. So, when using the Ward algorithm, Forest and Heat can detect clusters only if the community structure is distinct and there are almost no edges between clusters.

Let's now analyze the common features for all the measures, by which we can assess the impact of network topology on the Ward algorithm itself.

Due to the computation limits, we used networks with $n \leq 1000$ for clustering with the Ward method. However, even for this network size interval, in Figure \ref{fig:ward-all-a} we can see that the performance of the Ward method degrades when the network size increases.

Similarly to the Spectral method, the quality of clustering increases with the increase in the average degree (Figure \ref{fig:ward-all-b}) and decreases with the increase in $\mu$ (Figure \ref{fig:ward-all-f}). Also, according to Figure \ref{fig:ward-all:e}, many small clusters are better than few big clusters for the Ward method. The explanation for these properties is the same as for the Spectral method. 

Figure \ref{fig:ward-all-d} shows that the power law exponent for the community size distribution still doesn't essentially affect the efficiency of community detection, although there are more fluctuations in comparison to the Spectral method. According to Figure \ref{fig:ward-all-c}, the relation between the efficiency and $\tau_1$ is fuzzy, and it is hard to detect any common properties for all the measures.

We can also make some conclusions about the comparative efficiency of the Ward and the Spectral algorithms. According to the results of the experiments, the Spectral method outperforms the Ward method in most cases.

\section{Conclusion}
In this paper, we studied how the network topology affects the quality of community detection for such graph measures as Walk, Communicability, Forest, Heat, and PageRank. A variety of network topologies were generated using the LFR model, and resulting graphs were clustered using the Ward and the Spectral method in combination with each of the above measures.

As a result, we found that the efficiency of proximity measures depends on the network topology in some way. However, this dependence is not critical, and measures which are efficient for most topologies can be found. For the Spectral method, the most efficient measure is Walk. When the Ward method is used, the Walk and the Communicability measures outperform others in most cases.

Also, we have found some common features for all the measures. Using these common features, we can conclude how the algorithms themselves depend on the network topology. For example, the Ward and the Spectral methods prefer small clusters to big clusters.
\bibliography{measures_network_topology}
\end{document}